# Interstitially bridged van der Waals interface enabling stacking-fault-free, layer-by-layer epitaxy

*GunWoo Yoo[1,2,†], TaeJoon Mo[1,2,†], Yong-Sung Kim[3\*], Chang-Won Choi[1,4], Gunho Moon[1,4], Sumin Lee[1,4], Chan-Cuk Hwang[5], Woo-Ju Lee[1,2], Min-Yeong Choi[1,2], Jongyun Choi[1,4], Si-Young Choi[1,4,6,\*], Moon-Ho Jo[1,4,\*] and Cheol-Joo Kim[1,2,\*]*

[1]Center for Epitaxial van der Waals Quantum Solids, Institute for Basic Science (IBS), Pohang 37673, Republic of Korea

[2]Department of Chemical Engineering, Pohang University of Science and Technology (POSTECH), Pohang 37673, Republic of Korea

[3]Korea Research Institute of Standards and Science (KRISS), Daejeon 34113, Republic of Korea

[4]Department of Materials Science & Engineering, POSTECH, Pohang 37673, Republic of Korea

[5]Beamline Research Division, Pohang Accelerator Laboratory, Pohang 37673, Republic of Korea

[6]Department of Semiconductor Engineering, POSTECH, Pohang 37673, Republic of Korea

[†]These authors contributed equally: GunWoo Yoo, TaeJoon Mo

[\*]Correspondence to: yongsung.kim@kriss.re.kr, youngchoi@postech.ac.kr, mhjo@postech.ac.kr, kimcj@postech.ac.kr

**ABSTRACT**

Van der Waals (vdW) crystals are prone to twisting, sliding, and buckling due to inherently weak interlayer interactions. While thickness-controlled vdW structures have attracted considerable attention as ultrathin semiconducting channels, the deterministic synthesis of stacking-fault-free multilayers remains a persistent challenge. Here, we report the epitaxial growth of single-crystalline hexagonal bilayer $MoS_2$, enabled by the incorporation of Mo interstitials between layers during layer-by-layer deposition. The resulting bilayers exhibit exceptional structural robustness, maintaining their crystallinity and suppressing both rotational and translational interlayer misalignments even after transfer processes. Atomic-resolution analysis reveals that the Mo interstitials are located at a single sublattice site within the hexagonal lattice, where they form tetrahedral bonds with sulfur atoms from both $MoS_2$ layers, effectively anchoring the interlayer registry. Density functional theory calculations further indicate that these Mo atoms act as nucleation centers, promoting the selective formation of the hexagonal bilayer phase. This approach offers a robust strategy for the deterministic growth of multilayer vdW crystals with precisely controlled stacking order and enhanced interlayer coupling.

Deterministic growth of single-phase layered materials is essential for realizing phase-sensitive properties with uniformity and scalability. While thickness-controlled van der Waals (vdW) structures have garnered significant interest as ultrathin semiconducting channels[1,2] many layered compounds can adopt multiple polytypes through variations in interlayer stacking sequences, leading to distinct crystal symmetries and emergent physical properties[3,4]. Parameters such as temperature[5,6,7], pressure[8,9,10] and chemical environment[11,12,13] can thermodynamically bias the formation of specific polymorphs within a given phase diagram. However, the free energy differences between stable vdW polytypes—typically on the order of a few meV per unit cell—are exceedingly small due to weak interlayer interactions, making precise phase selection intrinsically challenging (see Fig. 1a for a schematic of bilayer transition metal dichalcogenides)[14,15,16].

Recent advances have demonstrated the controlled synthesis of multilayer transition metal dichalcogenides (TMDs) with rhombohedral stacking order, achieved via edge-assisted nucleation[17] and interfacial epitaxy[18]. These approaches leverage stronger interfacial interactions—on the order of ~100 meV per unit cell—between the TMDs and substrate step edges to stabilize a specific stacking configuration. Yet, the weak intrinsic interlayer bonding remains insufficient to resist shear or tensile deformation, leading to post-growth stacking variations such as phase coexistence with opposite out-of-plane polarities[19,20] and the formation of strained solitons[21,22,23]. This underscores the continued difficulty in achieving structurally robust, single-phase multilayers with deterministically engineered stacking.

Interstitial atom engineering has emerged as a promising strategy to overcome this limitation by strengthening interlayer coupling through the deliberate incorporation of atoms into the vdW gap. Theoretical studies suggest that specific transition metal interstitials—such as Mo, W, or Ta—can reside at high-symmetry sites between TMD layers, forming tetrahedral or octahedral covalent bonds with adjacent chalcogen atoms (S, Se, or Te)[24,25]. These covalent interlayer bridges can significantly raise the interlayer interaction energy—by up to hundreds of meV per unit cell—thereby stabilizing selected stacking configurations (e.g., 2H vs. 3R). This enhanced coupling provides a robust thermodynamic driving force for deterministic phase selection during growth.

Experimental realization of such covalently bonded layered materials has been achieved in several systems, most notably in Ta- and V-based TMDs under metal-rich growth conditions[12]. However, for technologically important semiconductors like $MoS_2$, intercalated phases are theoretically predicted to exhibit relatively high formation energies, and experimental demonstrations of stacking order control via interstitial incorporation remain limited. Thus, the development of interstitial-based strategies tailored for semiconducting TMDs such as $MoS_2$ represents a critical and largely unexplored frontier in the field.

**Results and Discussion**

**Layer-by-layer epitaxy of single crystalline bilayer $MoS_2$ film with interstitial Mo atoms**

Here, we develop a strategy to incorporate Mo interstitial atoms between layers for single-crystalline bilayer $MoS_2$ films. These interstitials function as atomic anchors, covalently bridging adjacent layers and promoting the formation of a uniform hexagonal phase with enhanced interlayer coupling (Fig. 1b). During layer-by-layer epitaxy, unidirectionally aligned monolayer $MoS_2$ grains coalesce into a continuous film on miscut c-plane sapphire via step-directed epitaxy[26] (steps (*i*) and (*ii*) in Fig. 1c; Fig. S1). Subsequent nucleation of monolayer grains on the first-grown $MoS_2$ surface leads to the formation of the second layer (steps (*iii*) and (*iv*) in Fig. 1c; Fig. S2). Notably, the growth was conducted under Mo-rich conditions, using a high-flux Mo supply with a partial vapor pressure ratio between Mo and S precursors, $P_{Mo}/P_S$ ~4×10$^{-2}$; See Methods). This condition was enabled by the high volatility of the metal-organic chemical vapor deposition (MOCVD) precursors[27].

High-angle annular dark-field scanning transmission electron microscopy (HAADF-STEM) was used to examine the atomic configurations of the sample grown under Mo-rich conditions (Fig. 1d). In the monolayer regions, Mo atoms exhibit strong scattering intensities arranged in a hexagonal lattice, while isolated sites with significantly enhanced intensities are observed. These sites exhibit an area density of $3 \times 10^{13}$ cm$^{-2}$, corresponding to approximately one interstitial per ~40 MoS$_2$ unit cells (Fig. S3). The intensity profile across these bright atomic sites aligns well with simulation data for a single Mo atom positioned on a Mo lattice site in monolayer MoS$_2$ (Fig. 1f, g). In bilayer regions, similarly bright atomic sites appear (Fig. 1e), vertically aligned with Mo sites in the bottom layer (Fig. 1h, i), comparable density of $2.4 \times 10^{13}$ cm$^{-2}$. In contrast, samples grown under Mo-poor conditions ($P_{Mo}/P_S$ ~2×10$^{-3}$; See Methods) show negligible evidence of interstitial or adatoms (Fig. S4).

We found that growth under Mo-rich conditions are essential for deterministic control of the stacking order. During the second-layer nucleation, Wulff-shaped hexagonal grains form with edges aligned perpendicular to the miscut direction of the sapphire substrate, as indicated by dotted lines in the dark-field transmission electron microscopy (DF-TEM) image (Fig. 2a), captured using an objective aperture to collect electrons diffracted by the ($\bar{1}$100) planes. The bilayer regions exhibit approximately four times the intensity of the monolayer regions, indicating hexagonal stacking order [28,29] (Fig. S5). Across the examined sample area, over 99% of bilayer domains (159 out of 160) exhibit hexagonal stacking order (Fig. S6), in stark contrast to samples grown under lower Mo concentrations, which show mixed hexagonal and rhombohedral phases (Fig. 2b; Fig. S7). Continued growth under Mo-rich conditions yields full bilayer coverage (step (*iv*)) with uniform diffraction patterns (Fig. 2c, d), confirming the deterministic formation of hexagonal stacking order.

Remarkably, bilayers grown under Mo-rich conditions maintain a single-phase structure without strained solitons, even after transfer of the samples, as shown in DF-TEM images taken from different diffraction spots (Fig. 2e, f). This structural robustness against local variations in stacking order induced by external shear or tensile strains suggests the presence of strong interlayer coupling. In contrast, samples grown under Mo-poor conditions display triangular grains on the first $MoS_2$ layer, oriented in opposite directions and exhibiting varying intensities in DF-TEM images acquired from an inner diffraction spot—corresponding to the hexagonal and rhombohedral phases, respectively (Fig. 2g). Additionally, strained solitons are abundantly observed in both phases in DF-TEM images taken from different diffraction spots (Fig. 2h), consistent with previous reports[21,22].

**Growth mechanism of single-crystalline bilayer $MoS_2$ with Mo interstitial**

To elucidate the mechanism determining the stacking order, we investigated the structural correlations between Mo interstitial atoms and interlayer stacking order using density functional theory (DFT) calculations. The two most stable atomic configurations of bilayer $MoS_2$ with a Mo interstitial both exhibit hexagonal stacking orders (Fig. 3a, b; Fig. S8). The lowest formation energy ($E_F$) is obtained when the Mo interstitial aligns vertically with the Mo atoms in both the top and bottom layers, occupying an octahedral site and forming covalent bonds with neighboring S atoms on each side, resulting in a total coordination number of six within the AB' stacking configuration (O-AB'; Fig. 3a). Within the AA' stacking configuration, the Mo interstitial can occupy a tetrahedral site, resulting in a total coordination number of four (T-AA'; Fig. 3b).

We compare the difference in $E_F$ ($\Delta E_F$) between T-AA' and O-AB' stackings as a function of lattice size ($N$) in the presence of an interstitial atom, as shown in Fig. 3c. The formation energies ($E_F$) of each structure are primarily governed by two contributions: $E_{Moi}$, representing the formation energy of a Mo interstitial and $E_{Bulk}$, representing the formation energy of the bulk bilayer. The positive $\Delta E_{Moi}$ for a Mo interstitial (gray data) indicates that the O-AB' stacking is energetically favored over T-AA' by 1.689 eV per Mo interstitial atom. In contrast, for the bilayer region separated by vdW gaps (orange data), $\Delta E_{Bulk}$ becomes negative, indicating that T-AA' is more favorable than O-AB'; this energy difference increases in magnitude as $N$ increases, corresponding to a larger bilayer area. The total $\Delta E_F$ changes sign at $N = 11$, beyond which the T-AA' configuration becomes thermodynamically most stable.

Based on these theoretical results, the proposed bilayer growth scenario is illustrated in Fig. 3d. Nucleation on top of the monolayer $MoS_2$ occurs with the assistance of a Mo interstitial, which reduces the activation barrier for nucleation. Initially, the structure adopts the O-AB' stacking order due to the small nucleus size. As the second-layer grain grows, the top grain translates by breaking two bonds with the Mo interstitial, resulting in the T-AA' configuration—consistent with atomic-resolution TEM observations (Fig. 1e). Although the experimentally confirmed average size for a single interstitial ($N \sim 7$) is somewhat smaller than the theoretically predicted threshold ($N = 11$) required to stabilize the T-AA' configuration, we hypothesize that additional Mo interstitials may be introduced following the formation of the AA'-stacked host. We note that the $\Delta E_{Moi}$ of at least 1.689 eV between different stacking orders is significantly higher than the $\Delta E_{Bulk}$ without interstitials (e.g., 1 meV per unit cell between hexagonal and rhombohedral phases; Fig. S9). Therefore, the interstitials can strongly direct the formation of a specific stacking order at the nucleation stage. Furthermore, they can suppress the formation of strained solitons, which would require breaking the strong covalent bonds between the interstitial and the MoS$_2$ surface. Previous theoretical studies have proposed using Mo interstitials to control stacking orders for enhancing interlayer interactions and tuning the stacking-dependent formation energy of bilayer phases[24,30,31]. However, to the best of our knowledge, this represents the first experimental demonstration of the deterministic formation of a specific interstitial and its utilization to control the stacking order.

**Structural and chemical properties of bilayer MoS$_2$ with Mo interstitials**

We further compare the structural and chemical properties of bilayer MoS$_2$ with and without interstitials. In the Raman spectra (Fig. 4a), the sample grown by MOCVD under Mo-poor conditions displays a single peak for the in-plane $E_{2g}^1$ vibration mode, whereas the as-grown sample synthesized under Mo-rich conditions exhibits two clearly distinct peaks. This peak splitting persists after transferring the film from the growth substrate (Fig. S10a), indicating that the splitting is not induced by substrate-film interactions. Similar peak splitting has been previously reported in bilayer MoS$_2$ containing covalently bonded Li interstitials[32]. We assign the peaks with lower and higher Raman shifts to $E_{2g}^1(-)$ and $E_{2g}^1(+)$, respectively. The $E_{2g}^1(-)$ mode, which exhibits stronger intensity, is red-shifted by 5.7 cm$^{-1}$ compared to the reported value for unstrained bilayer MoS$_2$[33,34]. Prior studies have shown that tensile strain can soften the $E_{2g}^1$ phonon mode[35,36], whereby the observed red-shift corresponds to a high tensile strain of approximately 1.6%. Such a high in-plane strain is unlikely to be maintained by weak vdW interactions with the substrate due to the intrinsic stiffness of the MoS$_2$ lattice. However, strong covalent bonding with interstitial atoms could plausibly induce and sustain this level of strain within the MoS$_2$ lattice.

X-ray photoelectron spectra (XPS) also reveal distinct differences between the two types of bilayer MoS$_2$. While peaks corresponding to Mo in the +4 oxidation state and S in the

−2 oxidation state dominate both spectra[37], the sample containing Mo interstitials exhibits symmetric broadening of the S 2p peak, resulting in an approximately 35% increase in the full width at half maximum (Fig. S10b). In contrast, the Mo 3d peak exhibits asymmetric broadening toward the lower binding energy, with additional shoulder peaks, attributed to Mo in lower oxidation states (red- and blue – highlighted peaks in Fig. 4b). These extra peaks are shifted by approximately 1 eV to lower binding energy relative to the main Mo 3d peak, which is significantly larger than the typical shift of ~0.5 eV associated with sulfur vacancies[38]. This observation supports our structural model of the bilayer with Mo interstitials, wherein excess Mo cations withdraw electrons from covalently bonded sulfur atoms, resulting in neighboring Mo atoms (highlighted as purple spheres in Fig. 4c) being less oxidized. The fraction of less-oxidized Mo atoms is approximately 11.5% of the total Mo content (based on the Mo $3d_{5/2}$ in the XPS spectrum), which is consistent with the STEM results (~11.3%).

We further note that the interstitials connecting adjacent layers to guide the stacking order exhibit identical atomic configurations, including their orientations. A Mo interstitial with tetrahedral bonding configuration can align with Mo atoms in either the bottom or top layer; however, TEM characterization (Fig. 1e) shows that all observed interstitials align with Mo atoms in the bottom layer. The interstitial bonds, with a fixed out-of-plane polarity, are expected to be kinetically driven during layer-by-layer epitaxy. When the octahedral Mo configuration transitions to a tetrahedral one (Fig. 3d), it is energetically favorable to slide the smaller top $MoS_2$ layer rather than the bottom layer by breaking the bonds between the Mo interstitial and the top layer.

**Electronic properties of single-crystalline bilayer $MoS_2$**

The robust stacking order and interstitial bonds with fixed bonding polarity result in uniform interlayer interactions across the sample. To investigate the electronic band structure, we performed angle-resolved photoemission spectroscopy (ARPES) on the bilayer $MoS_2$ samples. The ARPES measurement was conducted over a characterization area of approximately $300 \times 150$ μm², spanning a film region formed by the merging of multiple nuclei. Despite this, the bilayer $MoS_2$ grown under Mo-rich conditions exhibits single-crystalline band structures, as evidenced by the contour map of photoelectron intensity across momentum space at various binding energies (Fig. S11a, b). The valence band dispersion along the high-symmetry Γ–K direction (Fig. S11c) was extracted from the extrema of the photoelectron intensity (red line in Fig. S11a) and aligns closely with previously reported data for a single crystal of hexagonal bilayer $MoS_2$ (dotted black line)[39].

The bilayer grown under Mo-poor conditions also exhibits characteristic features of a bilayer structure, with band extrema located at similar energy levels along the high-symmetry Γ–K–M path (Fig. S11d). However, the intensity profiles at the Γ point, as a function of binding energy across the valence band maxima, reveal distinct differences between the two samples (Fig. 5a-b). Theoretically, a bilayer crystal exhibits two distinct valence bands at the Γ point near the energy maximum[40,41], denoted $Γ^{1st}$ and $Γ^{2nd}$, which are separated by approximately 0.71 eV. In the sample grown under Mo-rich conditions, the $Γ^{1st}$ valence band maximum appears as a clear intensity peak (Fig. 5c, red plot). In contrast, the sample grown under Mo-poor conditions exhibits significantly weaker intensity at the $Γ^{1st}$ band (Fig. 5c, blue plot). Additionally, the Mo-poor sample shows an extra spectral peak, indicated by a green arrow in Fig. 5c, which does not correspond to any theoretically predicted band in either hexagonal AA' or rhombohedral AB/BA stacking configurations of stable bilayer $MoS_2$ (Fig. 5c, right panel)[42].

The $\Gamma^{1st}$ band, associated with hybridized states between interfacing layers, strongly depends on out-of-plane interlayer interactions[41,42,43,44,45,46]. Efficient and uniform electronic coupling between layers in the Mo-rich sample likely accounts for the distinct peak at $\Gamma^{1st}$. However, local regions with stacking faults—such as strained solitons abundantly observed in Fig. 2h—may lead to weak hybridization of interlayer states, shifting the $\Gamma^{1st}$ band closer to the $\Gamma^{2nd}$ and toward lower binding energy. Notably, the energy of the newly emerged band aligns with the theoretically predicted position of the $\Gamma^{1st}$ band in the thermodynamically unstable AA stacking configuration, which appears in regions of stacking faults, such as strained solitons, and exhibits weak interlayer interactions[42,44].

We also investigate electronic transport across stacking-fault-free, single-crystalline bilayer $MoS_2$ in a field-effect transistor (FET). Atomically thin $MoS_2$ has gained attention as a channel material for low-power electronics[2]. While previously reported epitaxial growths of multilayer $MoS_2$ with aligned crystallinity[14] has shown enhanced carrier mobility by minimizing the carrier scattering at tilted grain boundaries, abundant stacking faults exist in the multilayers with conductive twin boundaries[47], which can increase off-state currents. To characterize the intrinsic properties of the $MoS_2$ sample while minimizing external disorders[48], single-crystalline hexagonal boron nitride (hBN) and graphite—exfoliated from bulk crystals—are employed as dielectric and gate materials, respectively. In the transconductance curve as a function of gate voltage (Fig. 5e, f), the electron field-effect mobility is estimated to be approximately 35 cm²/V·s. Notably, the on/off ratio approaches $10^9$, which is among the highest reported for multilayer $MoS_2$ channels (Fig. 5g). The device also maintains stable electrical performance during extended operation (Fig. S12). In a direct comparison of electrical properties between devices with the same geometry, FETs fabricated from bilayer $MoS_2$ channels grown under Mo-poor conditions exhibit significantly degraded performance (Fig. S13). While various factors may contribute to this degradation, stacking variations within the channel are considered the primary factor determining the performance differences (see Fig. S13 and 14 for discussion).

**Conclusions**

In summary, we present a method for synthesizing large-area, single-crystalline vdW multilayers that resist sliding or twisting under mechanical stress. By introducing single-atom interstitials at the interlayer during layer-by-layer epitaxy—under precisely controlled chemical environments in MOCVD—we enable the formation of dilute atomic anchors that connect layers in the thermodynamically most stable configuration. This approach guides stacking order with enhanced mechanical robustness and improved performance in electronic transport.

While the current study primarily focuses on the growth of bilayer $MoS_2$, the interstitial engineering strategy is applicable to multilayer systems beyond bilayers, where the chemical environment during growth is controlled to deterministically form specific types of atomic sites that guide the interlayer stacking order (Fig. S15, S16). Furthermore, this interstitial strategy for controlling stacking order can be extended to other TMDs with identical outer electron configurations of their constituent elements, supporting the general validity of the proposed growth mechanism (Fig. 3, Fig. S17). Finally, our strategy is adaptable to various vdW crystals incorporating different interstitial elements, enabling deterministic control of interlayer interactions and crystal symmetry.

**Supporting information.**

Epitaxial growth of monolayer $MoS_2$, atomic-resolution images and phase distribution analysis of bilayer $MoS_2$ grown under Mo-rich and Mo-poor conditions, calculated formation energies of bilayer $MoS_2$ structures containing Mo interstitials, Raman and X-ray photoelectron spectroscopy results of bilayer $MoS_2$, electronic band dispersions of bilayer $MoS_2$, electronic transport characteristics of bilayer $MoS_2$, stacking order analysis in trilayer $MoS_2$, and transmission electron microscopy analysis of bilayer $WS_2$ grown under W-rich conditions.

**Methods**

TMDs growth

TMDs (MoS$_2$, WS$_2$) films were grown using a home-built MOCVD system. Sapphire substrates with a polished C-plane <0001> surface, miscut at 15° along the A-axis of <1120>, were used as epi-substrates. Before growth, the substrates were annealed at 1000 °C for 3 h in an Ar/H$_2$ atmosphere to induce surface reconstruction with atomic steps. During growth, NaCl and KCl powder were introduced near the furnace inlet to act as a growth promoter, respectively the case of MoS$_2$ and WS$_2$, facilitating precursor injection.

Mo(CO)$_6$ (Sigma-Aldrich 577766, ≥ 99.9%), W(CO)$_6$ (Sigma-Aldrich 472956, ≥ 99.9%), and (C$_2$H$_5$)$_2$S (Sigma-Aldrich 107247, 98%) served as the Mo, W and S precursors, respectively. Each precursor was housed in a separate chamber, and their vapors were carried into the growth chamber using Ar as the carrier gas. Growth was conducted at 810 ˚C under a mixture of Ar and H$_2$ flows.

Under Mo-rich conditions, the flow rates for Ar, H$_2$, (C$_2$H$_5$)$_2$S with Ar carrier gas and Mo(CO)$_6$ with Ar carrier gas were 500, 2, 0.1, and 2.0 sccm, respectively. For Mo-poor conditions, the corresponding flow rates were 90, 4, 0.4, and 0.4 sccm. The partial pressures of the Mo and S precursors in the growth chamber, $P_{Mo}$ and $P_S$, were estimated based on the vapor pressures in their respective precursor chambers and the carrier gas flow rates. For the Mo-rich condition, $P_{Mo} = 8.5 \times 10^{-5}$ Torr and $P_S = 2.1 \times 10^{-3}$ Torr; for the Mo-poor condition, $P_{Mo} = 1.7 \times 10^{-5}$ Torr, $P_S = 8.6 \times 10^{-3}$ Torr.

Under W-rich conditions, the flow rates for Ar, H$_2$, (C$_2$H$_5$)$_2$S with Ar carrier gas and W(CO)$_6$ with Ar carrier gas were 700, 2.5, 0.1, and 3.0 sccm, respectively. For W-poor conditions, the corresponding flow rates were 80, 0, 0.2, and 0.3 sccm. The partial pressures of the W and S precursors in the growth chamber, $P_W$ and $P_S$, were estimated based on the vapor pressures in their respective precursor chambers and the carrier gas flow rates. For the W-rich condition, $P_W = 8.6 \times 10^{-5}$ Torr and $P_S = 2.1 \times 10^{-3}$ Torr; for the W-poor condition, $P_W = 8.6 \times 10^{-6}$ Torr, $P_S = 4.2 \times 10^{-3}$ Torr.

DFT calculations

DFT calculations were performed as implemented in the Vienna *ab initio* simulation package (VASP) code[55]. The projector-augmented wave (PAW) pseudopotentials[56,57] and a kinetic energy cutoff of 500 eV were applied. The Perdew, Burke, and Ernzerhof (PBE) functional within the generalized gradient approximation (GGA) was used for the exchange correlation functional[58], and the vdW-D3 correction proposed by Grimme was used to describe the long-range vdW interaction[59]. Spin-polarization was included. The calculated hexagonal lattice constant of bulk MoS$_2$ was 3.161 Å. The in-plane 6×6 and out-of-plane 4-MoS$_2$-layer-thick supercell of MoS$_2$ was used, and the Γ, K-point was used for the Brillouin zone summation. The local atomic configurations were relaxed to less than 0.001 eV/Å in the Hellmann–Feynman forces. In calculations of formation energies, the total energies of a Mo bcc metal and a S orthorhombic crystal were chosen for the Mo- and S-rich limits, respectively.

TEM

For DF-TEM imaging from selected diffraction spots with the objective filtering aperture, $MoS_2$, $WS_2$ samples were transferred onto a 20 nm thick silicon nitride membrane, then imaging was conducted using a JEOL JEM-2100F operated at 200 kV.

For STEM imaging, samples were transferred onto a Quantifoil TEM Cu-grid. The STEM data were obtained using a JEOL JEM-ARM 200F equipped with a fifth-order spherical aberration corrector (ASCOR, CEOS GMBH) at the Materials Imaging & Analysis Center of POSTECH in South Korea. The acceleration voltage was set to 80 kV, and a high-angle annular dark-field detector (inner angle: 54 mrad, outer angle: 216 mrad) with 40-μm condenser lens aperture was used to acquire the STEM images.

ARPES

ARPES measurements were conducted at the 10D beamline of the Pohang Accelerator Laboratory (PAL) using a Scienta DA30 analyzer under ultra-high vacuum (~$10^{-9}$ Torr) at room temperature, with a photon energy of 56 eV. Samples were prepared by depositing an 80 nm thick Au layer onto the as-grown $MoS_2$, serving as a transfer medium. The Au-coated surface was mounted onto the sample holder and secured with Ta foil. In the load-lock chamber, once ultra-high vacuum was achieved, the sapphire substrate was detached from the $MoS_2$ to avoid contamination prior to measurement. The Fermi level was calibrated using an Au reference.

Device fabrication and transport measurements

For back-gated field-effect transistors (FETs) with an hBN dielectric (Fig. 5d, S12), graphite (~10 nm thick) and hBN (30 nm thick) were sequentially exfoliated from bulk crystals and transferred onto a $Si/SiO_2$ substrate. The as-grown $MoS_2$ was then transferred onto the graphite/hBN stack using a PMMA superlayer. Electron-beam lithography was employed to define the electrode and channel regions, followed by deposition of Bi/Au (15 nm/45 nm) to form source and drain contacts and Ti/Au (5 nm/ 40 nm) was deposited on the pad region.

For back-gated FETs with an $Al_2O_3$ dielectric (Fig. S13), Cr/Au (10 nm/40 nm) was first deposited on a $Si/SiO_2$ substrate. A 30 nm thick $Al_2O_3$ layer was subsequently deposited via atomic layer deposition (Plus 200, QUROS). The $MoS_2$ film was transferred onto the $Al_2O_3$ surface, and device geometry was defined using photolithography. Source and drain contacts were formed with Bi/Au (15 nm/45 nm), and $MoS_2$ channels were patterned using oxygen-reactive ion etching. Electrical measurements were conducted under vacuum (~$10^{-5}$ Torr) at room temperature.

Other characterizations

Raman and photoluminescence spectroscopy were performed at room temperature using a 532 nm laser (power: 10–30 mW), focused onto the sample through a 100× microscope objective lens (0.9 N.A.). The spectrometer was configured with a grating of 1800 grooves/mm.

X-ray photoelectron spectroscopy (XPS) was carried out using a Thermo Fisher Scientific system equipped with an Al Kα X-ray source (hν = 1486.6 eV, 15 kV, 100 W), producing a spot size of 400 μm in diameter. Charge compensation was applied using dual flood guns (low-energy electrons and $Ar^+$ ions).

# Figures

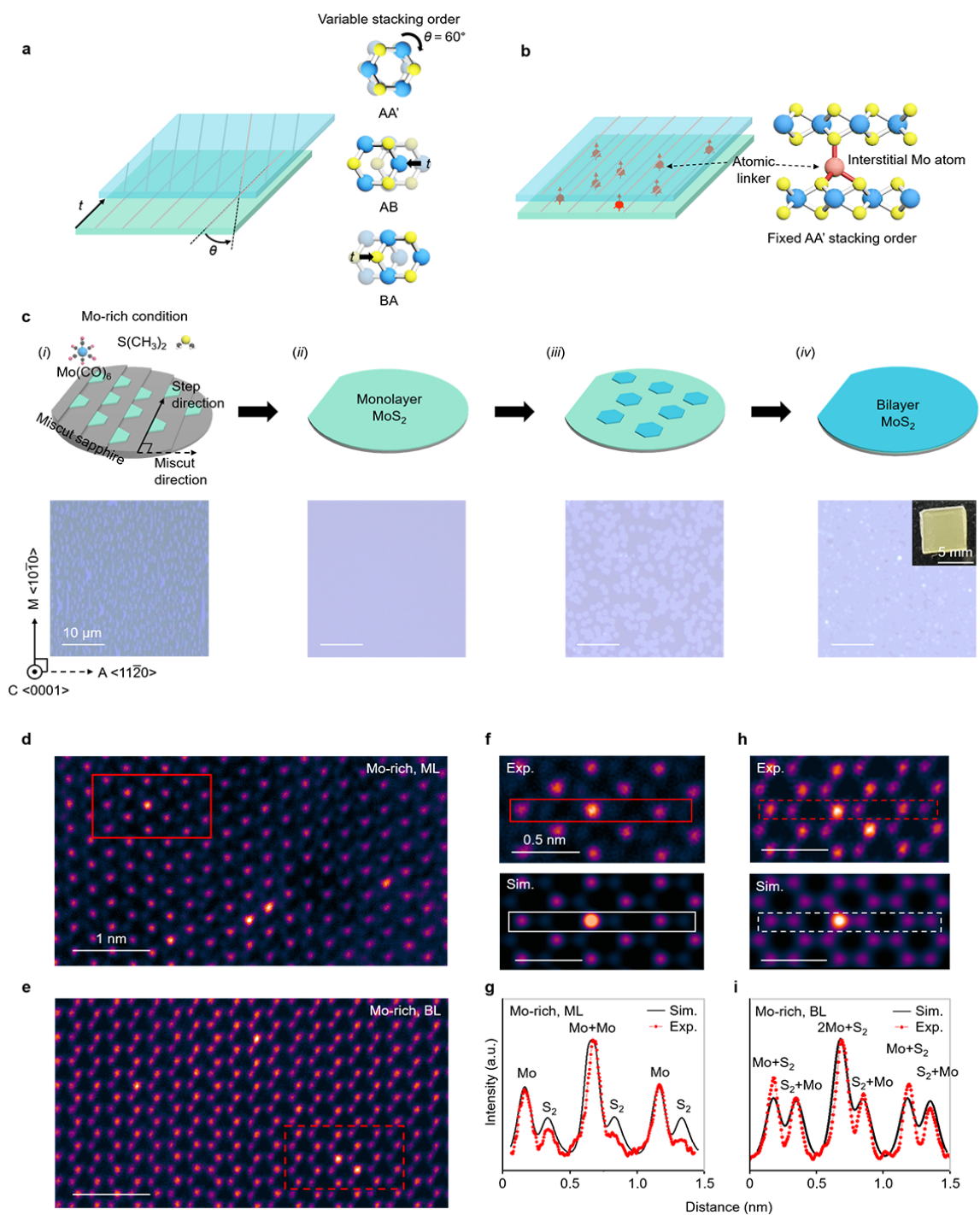

**Fig. 1. Layer-by-layer epitaxy of MoS$_2$ with interstitial single Mo atoms.**

(a) Schematic illustrations of bilayer MoS$_2$ with various stacking orders induced by layer translation and rotation. (b) Schematic of bilayer MoS$_2$ incorporating Mo interstitial atoms, which act as atomic linkers covalently connecting adjacent layers. (c) Schematic of the layer-by-layer epitaxy process of bilayer MoS$_2$ on a miscut sapphire substrate, accompanied by optical reflectance images at each stage. (d), (e) STEM images of monolayer (d) and bilayer (e)

MoS$_2$ grown under Mo-rich conditions. (f) Experimental STEM image from the region highlighted in (d) (top) and a corresponding simulated image of a monolayer containing a Mo adatom located at a Mo lattice site (bottom). (g) Intensity profiles extracted from the boxed region in (f), comparing experimental (red) and simulated (black) data. (h) Experimental STEM image from the region highlighted in e (top) and the corresponding simulated image of a bilayer with a Mo adatom at a Mo site in the lower monolayer (bottom). (i) Intensity profiles extracted from the boxed region in (h), showing experimental (red) and simulated (black) results.

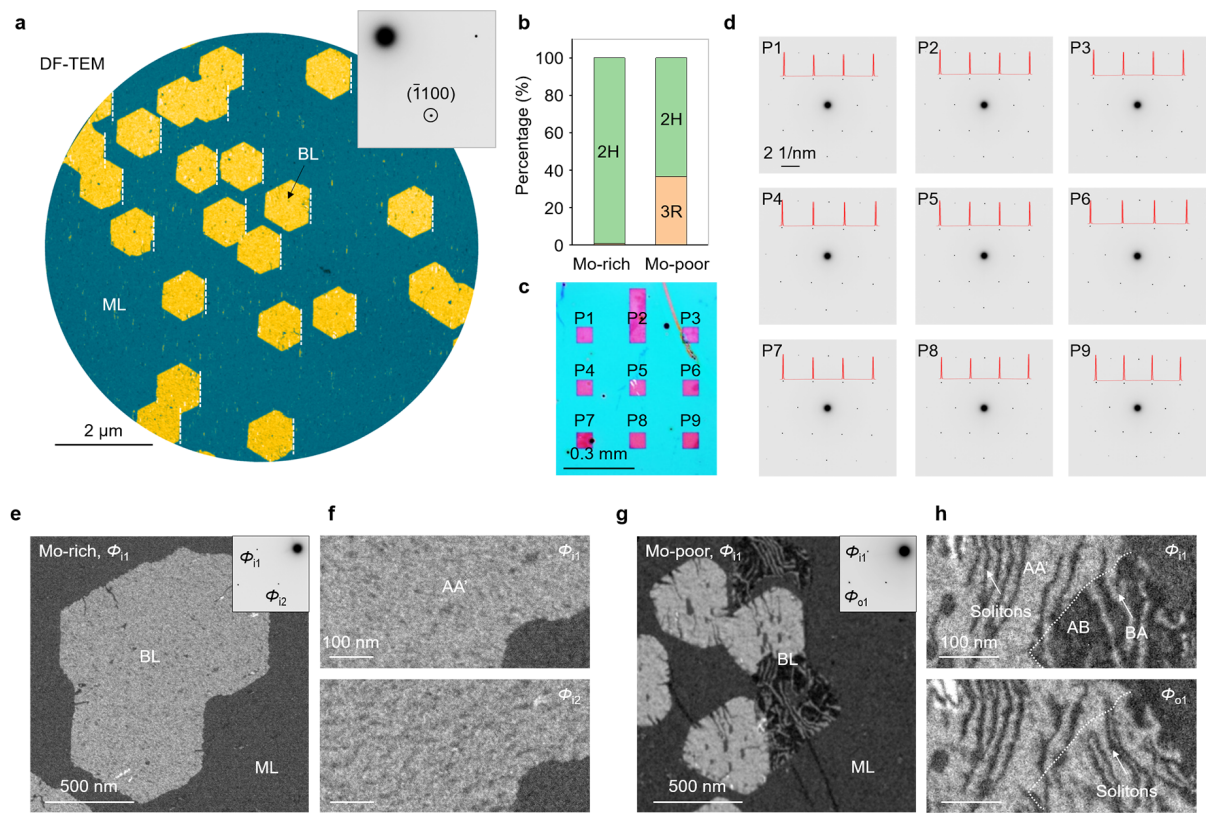

**Fig. 2. Uniform growth of robust, single-phase bilayer MoS$_2$ film.**

(a) DF-TEM image obtained using an inner diffraction spot (upper right). The yellow hexagonal grain corresponds to the bilayer MoS$_2$ region, while the cyan background indicates the monolayer region. (b) Histogram comparing the bilayer phase distributions grown under Mo-rich and Mo-poor conditions. (c) Optical reflectance image of bilayer MoS$_2$ grown under Mo-rich conditions and transferred onto a silicon nitride TEM grid. Positions P1–P9 denote suspended regions above the grid windows. (d) Selected-area electron diffraction (SAED) patterns acquired from the positions marked in (c). The red line indicates the intensity profile across the diffraction spots. (e) DF-TEM image of a bilayer grain grown under Mo-rich conditions; the corresponding diffraction pattern is shown in the upper right. (f) DF-TEM images obtained using different diffraction spots from (e). (g) DF-TEM image of bilayer grains grown under Mo-poor conditions; corresponding diffraction pattern shown in the upper right. (h) DF-TEM images obtained using an inner (top) and outer (bottom) diffraction spot.

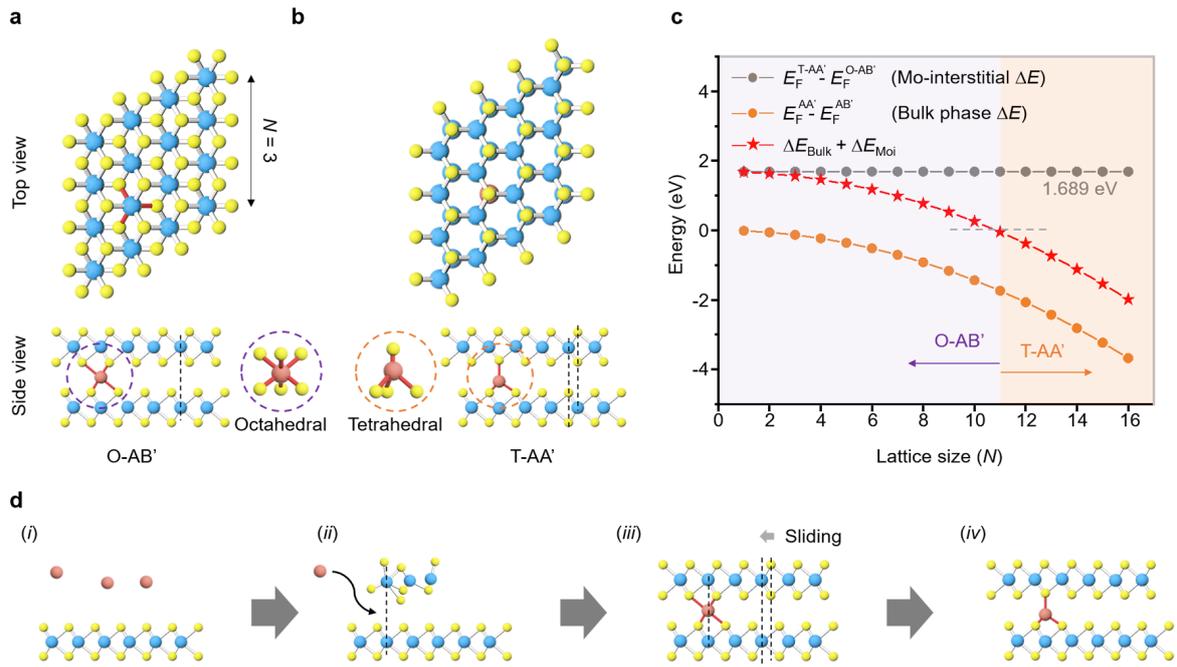

**Fig. 3. Mechanism of single-adatom-seed-driven epitaxy of hexagonal bilayers.**

(a), (b) Top and side views of the two most stable atomic configurations of hexagonal bilayer MoS$_2$ with a Mo interstitial, based on DFT calculations. $N$ denotes the number of MoS$_2$ unit cells along one axis of the simulation cell. Red spheres indicate Mo interstitials. (c) Formation energy as a function of $N$. Grey: energy difference of Mo interstitials between T-AA' and O-AB' configurations; orange: energy difference between bulk AA' and AB' phases; red: total energy difference between T-AA' and O-AB', combining interstitial and bulk contributions. (d) Schematic illustration of the proposed layer-by-layer epitaxy mechanism for hexagonal bilayer MoS$_2$ incorporating a Mo interstitial as a nucleation seed.

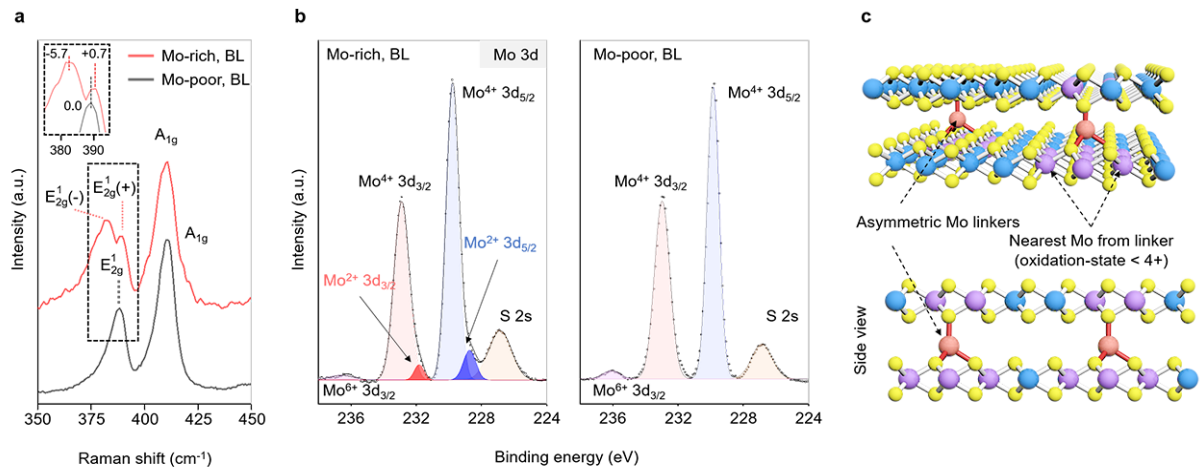

**Fig. 4. Structural and chemical properties of bilayer MoS$_2$ with Mo interstitials.**

(a) Raman spectra of bilayer MoS$_2$ grown under Mo-rich (red) and Mo-poor (black) conditions. The Mo-rich sample exhibits peak splitting, referenced to the $E_{2g}^1$ mode of the Mo-poor sample (inset). (b) XPS spectra of bilayer MoS$_2$ grown under Mo-rich (left) and Mo-poor (right) conditions, showing the Mo 3d. S 2p peaks are shown in Fig. S10. (c) Schematic illustration of bilayer MoS$_2$ with a Mo interstitial acting as a linker between layers. Red spheres denote Mo interstitials; purple spheres represent adjacent Mo atoms, whose oxidation states are affected by the nearby interstitials.

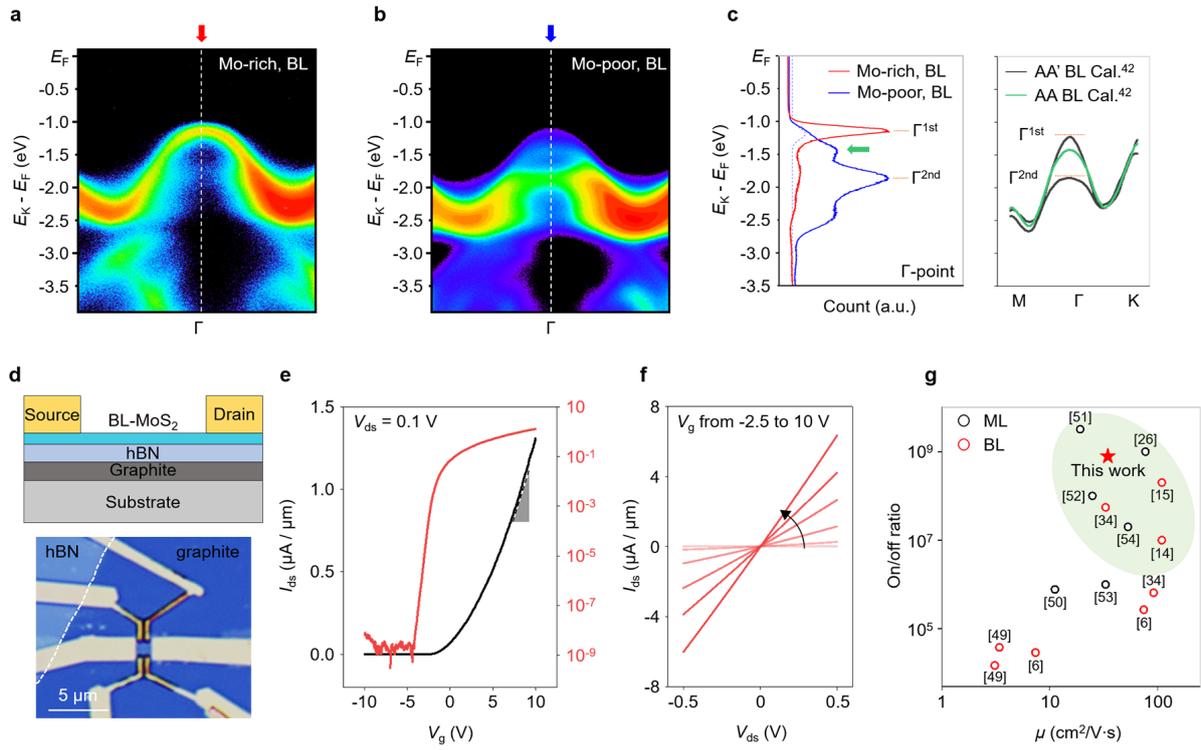

**Fig. 5. Band structure and electronic properties of bilayer MoS$_2$.**

(a), (b) Band dispersion of Γ direction obtained by ARPES which grown under Mo-rich and Mo-poor conditions, respectively. (c) Photoelectron intensity as a function of energy at the Γ point. Left: red and blue lines represent data from samples grown under Mo-rich and Mo-poor conditions, respectively. Right: calculated electronic band dispersions of 2H bilayer MoS$_2$ with AA' (black) and AA (green) stacking orders, adapted from reference[42]. (d) Schematic of the device structure (top) and optical microscopy image of the fabricated field-effect transistor (FET) device (bottom). Scale bar, 5 μm ($L_{ch}$ = 1.35 μm, $W_{ch}$ = 2.2 μm). (e) Transconductance curve ($I_{ds}$ vs $V_g$) measured at $V_{ds}$ = 0.1 V for bilayer MoS$_2$ grown under Mo-rich conditions. (f) Output characteristics ($I_{ds}$ vs $V_{ds}$) of the same device shown in (e). $V_g$ was varied from -2.5 V to 10 V in 2.5 V steps. (g) Benchmark comparison of CVD-grown mono- and bilayer MoS$_2$ in terms of on/off ratio and field-effect mobility, compiled from references[6, 14, 15,26,34,49-54].

**Funding sources**

This research was supported by the National R&D Program through the National Research Foundation of Korea (NRF) funded by the Ministry of Science and ICT (2023R1A2C2005427, RS-2023-00258309), and the Institute for Basic Science (IBS-R034-D1).

**Author contributions**

G.W.Y., T.J.M., M.-H.J., and C.-J.K. designed the project. G.W.Y., W.-J.L., and M.-Y.C. built MOCVD systems. G.W.Y. and T.J.M. synthesized $MoS_2$ samples. Y.-S.K. performed the DFT calculations. C.-W.C. and S.-Y.C. performed the STEM imaging and analysis. G.W.Y., G.H.M., S.M.L., and J.Y.C. fabricated the devices and conducted transport measurements. G.W.Y., T.J.M., and W.-J.L. performed ARPES measurements and C.-C.H. assisted to analyze the ARPES measurement data. G.W.Y., T.J.M., and C.-J.K. wrote the manuscript with inputs from all authors.

**Competing declarations**

The authors declare no conflict of interests.

# Supporting Information

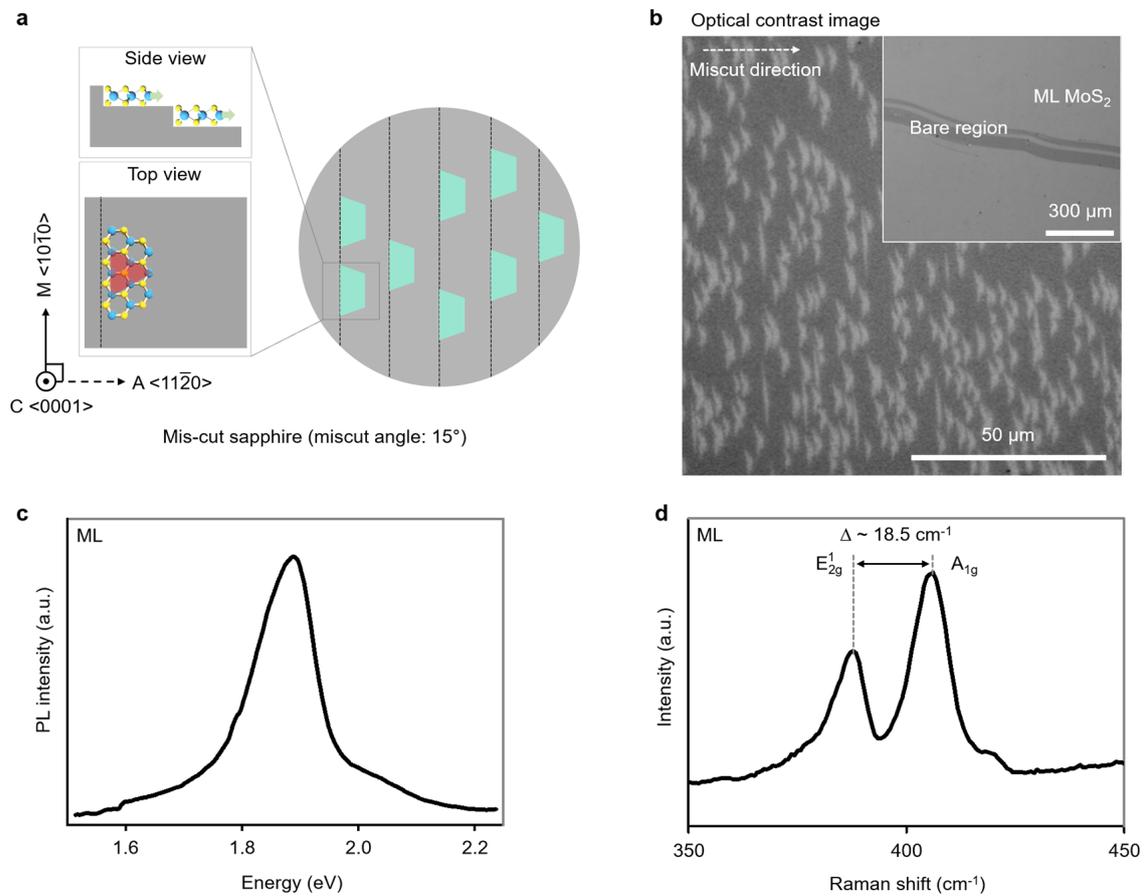

**Fig. S1. Unidirectional growth of monolayer MoS₂**

(a) Schematic illustration of edge-assisted unidirectional growth of monolayer MoS$_2$ on 15° miscut sapphire substrate. (b) Optical contrast image showing aligned monolayer grains with truncated triangular shapes oriented along the step direction. The upper panel highlights the coalescence of monolayer regions into a continuous film, which exhibits higher optical contrast than the bare sapphire substrate. (c), (d) Photoluminescence (PL) and Raman spectra of the monolayer MoS$_2$ film. The strong PL intensity at ~1.9 eV confirms the direct bandgap of monolayer MoS$_2$. The Raman spectrum shows two distinct peaks corresponding to the $E_{2g}^1$ and $A_{1g}$ vibrational modes of MoS$_2$, with a frequency difference of ~19 cm$^{-1}$, confirming the monolayer film.

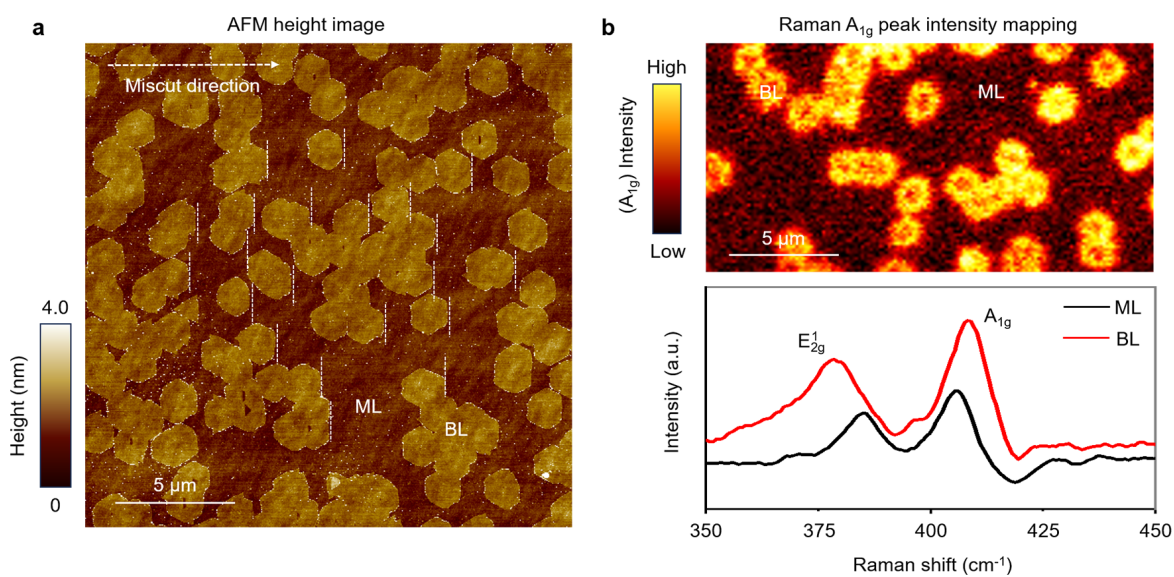

**Fig. S2. Unidirectional hexagonal partial grains of bilayer MoS$_2$**

(a) AFM height image of bilayer MoS$_2$ grains grown via layer-by-layer epitaxy. The second-layer grains exhibit a hexagonal shape and are aligned along the step direction. (b) Raman A$_{1g}$ intensity mapping and corresponding spectra for the monolayer and bilayer regions. The bilayer region shows approximately twice the Raman intensity compared to the monolayer region.

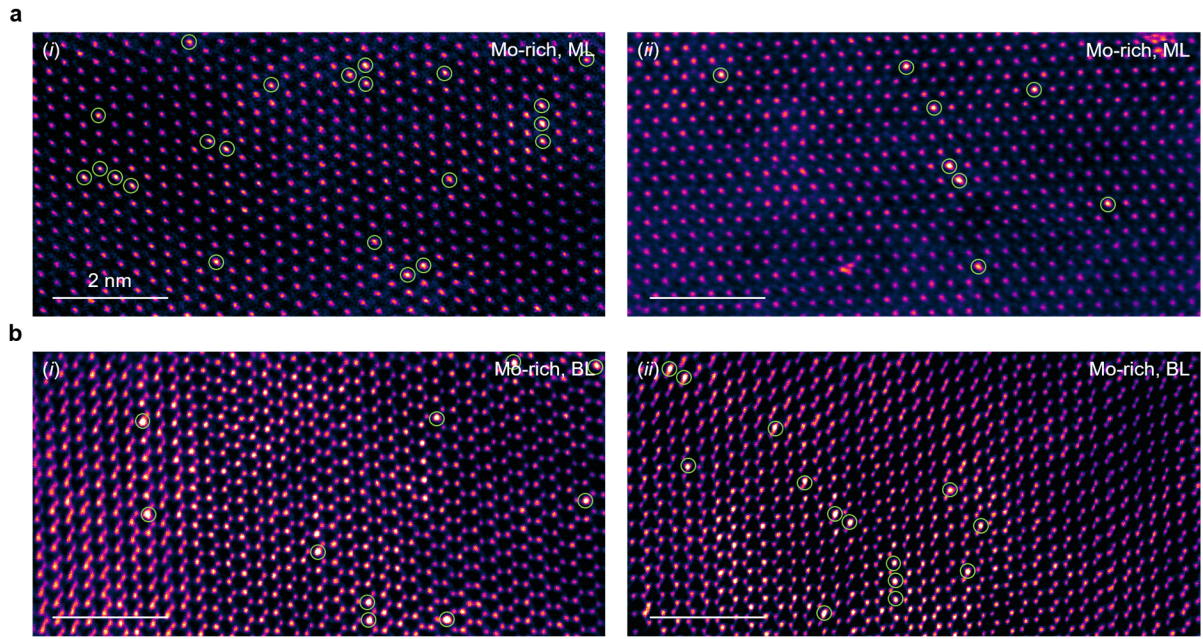

**Fig. S3. Selective adsorption of Mo adatoms under Mo-rich conditions.**

(a), (b) (*i*)–(*ii*) STEM images of monolayer and bilayer MoS$_2$ regions grown under Mo-rich conditions. Bright contrast features correspond to individual Mo adatoms, which are found to selectively occupy a specific sublattice site coinciding with the Mo positions in the first layer of the MoS$_2$ lattice. The average density of Mo adatoms is approximately $3 \times 10^{13}$ cm$^{-2}$ for the monolayer and $2.4 \times 10^{13}$ cm$^{-2}$ for the bilayer, corresponding to roughly one Mo adatom per 40 MoS$_2$ unit cells in the monolayer.

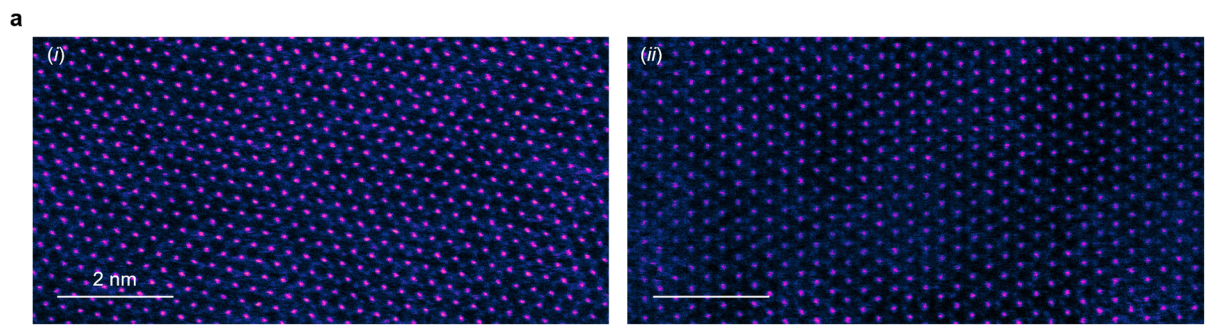

**Fig. S4. STEM image of monolayer MoS₂ grown under Mo-poor conditions.**

(a) (*i*)–(*ii*) STEM images of monolayer MoS$_2$ grown under Mo-poor conditions. Bright contrast features corresponding to individual Mo adatoms are absent.

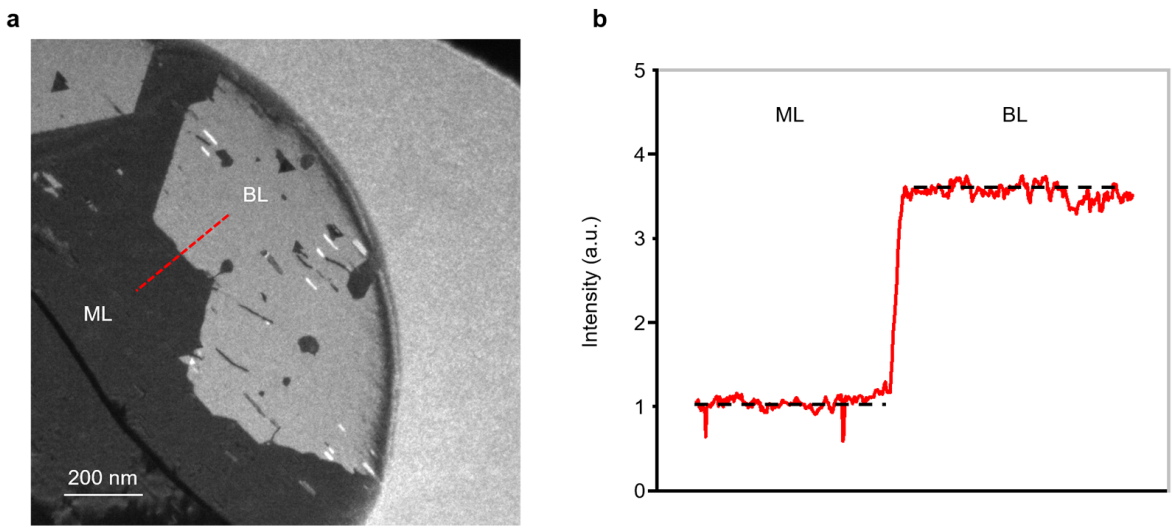

**Fig. S5. Confirmation of 2H stacking order using dark-field TEM.**

(a) Inner-spot dark-field TEM image of $MoS_2$ grown under Mo-rich conditions. (b) Line profile across the monolayer and bilayer regions. The bilayer region exhibits an intensity approximately four times higher than that of the monolayer, consistent with the 2H stacking order.

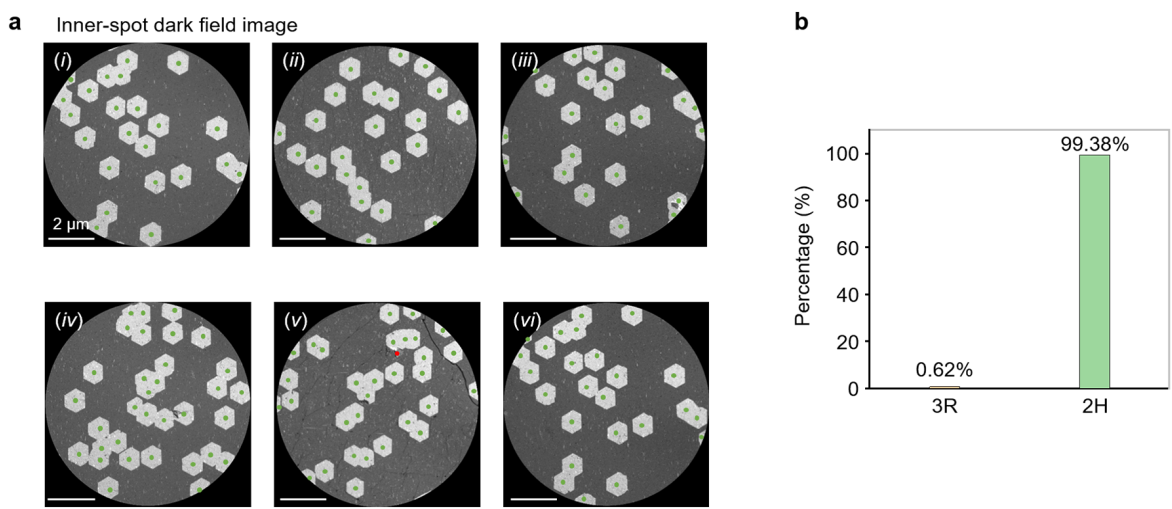

**Fig. S6. Phase distribution of bilayer grains grown under Mo-rich conditions.**

(a) (*i*)–(*vi*) Inner-spot dark-field TEM images of different regions of bilayer $MoS_2$ grown under Mo-rich conditions. The hexagonal grains exhibit uniform intensity approximately four times higher than that of the monolayer region, indicating the presence of the 2H phase. (b) Statistical comparison of bilayer grains exhibiting brighter contrast than the monolayer region (2H phase) with those exhibiting contrast similar to the monolayer (3R phase). Over 99% of the grains correspond to 2H-phase bilayer $MoS_2$.

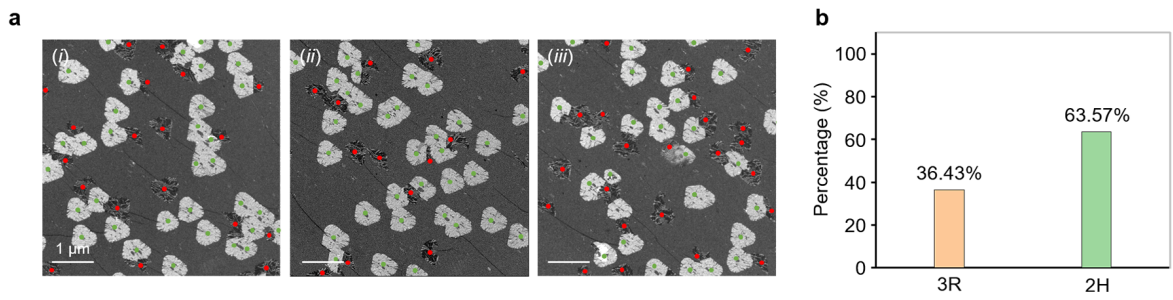

**Fig. S7. Phase distribution of bilayer grains grown under Mo-poor conditions.**

(a) (*i*)–(*iii*) Inner-spot dark-field TEM images of bilayer $MoS_2$ grown under Mo-poor conditions. Bright and dark triangular grains correspond to the 2H and 3R phases, respectively, with each type oriented in opposite directions. (b) Quantitative comparison of bilayer grain phases. Approximately 63.57% of the grains are in the 2H phase, while the remaining correspond to the 3R phase.

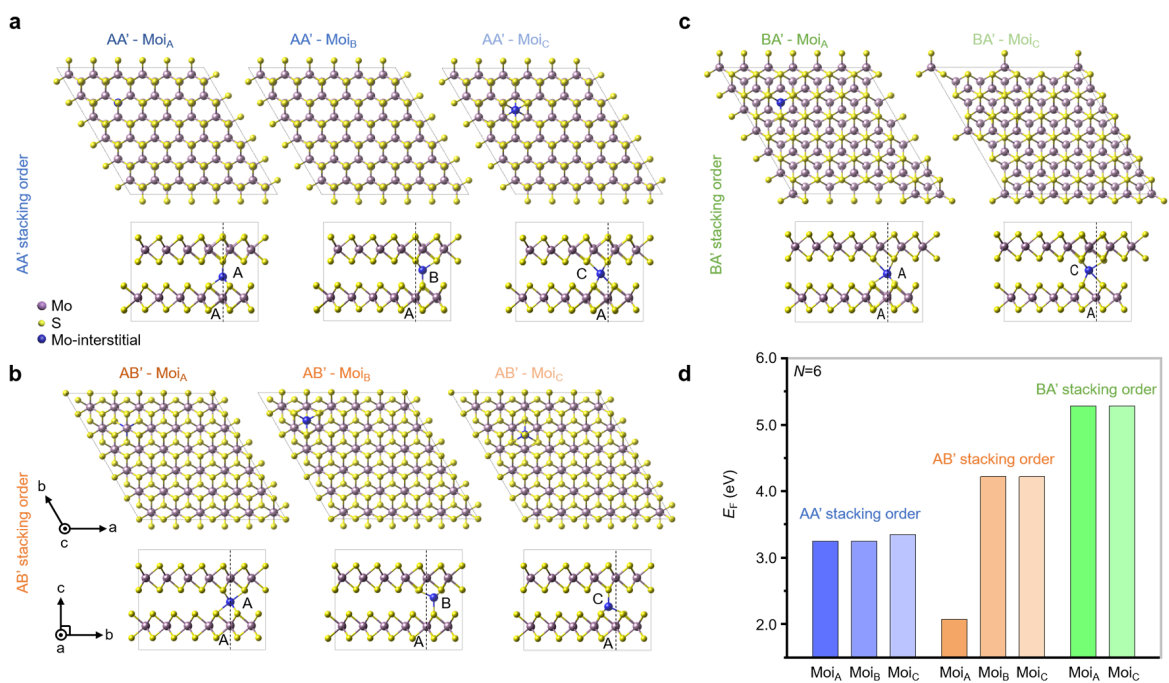

**Fig. S8. Formation energy of bilayer MoS$_2$ as a function of Mo interstitial site.**

(a)–(c) Atomic models of bilayer MoS$_2$ with Mo interstitials located at different sites: directly above a Mo atom in the first layer (Moi$_A$), above a sulfur atom (Moi$_B$), and at a hollow site (Moi$_C$). (d) Calculated formation energies for various stacking orders and Mo interstitial configurations. The BA'-Moi$_B$ configuration was too unstable for consideration. The most energetically favorable structures include AA stacking with Mo interstitials at Moi$_A$, Moi$_B$, and Moi$_C$ sites, as well as AB stacking with a Mo interstitial at the Moi$_A$ site. Although AA'-Moi$_B$ is the inversion counterpart of AA'-Moi$_A$ and thus exhibits the same formation energy under free-standing conditions, only the AA'-Moi$_A$ configuration is expected to form under the layer-by-layer epitaxy mode used in our experiments.

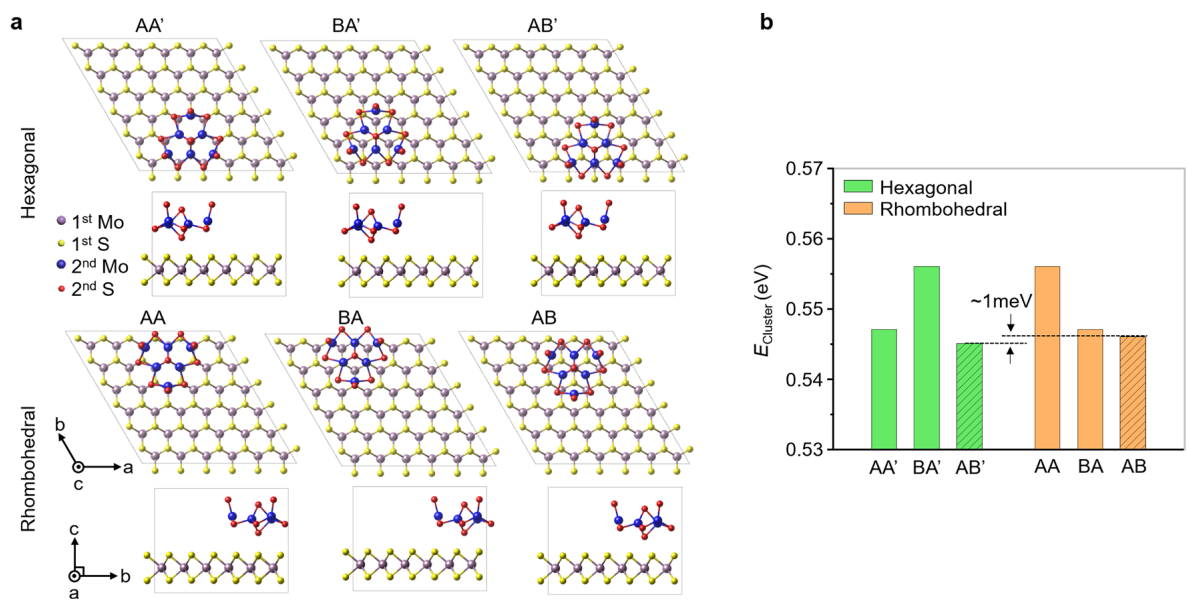

**Fig. S9. Cluster energy of various bilayer MoS$_2$ polytypes without interstitial.**

(a) Atomic models of possible bilayer MoS$_2$ polytypes under Mo-rich conditions ($\mu_s = -1.559$ eV). (b) Comparison of cluster energies for different bilayer MoS$_2$ polytypes. Among the hexagonal and rhombohedral phases, the AB and AB′ stacking configurations are the most stable, respectively, with a cluster energy difference of approximately 1 meV. Calculations were performed using an $N = 3$ triangular cluster with the S1-type sulfur edge configuration, which is the most stable edge structure under Mo-rich conditions ($\mu_s = -1.559$ eV).

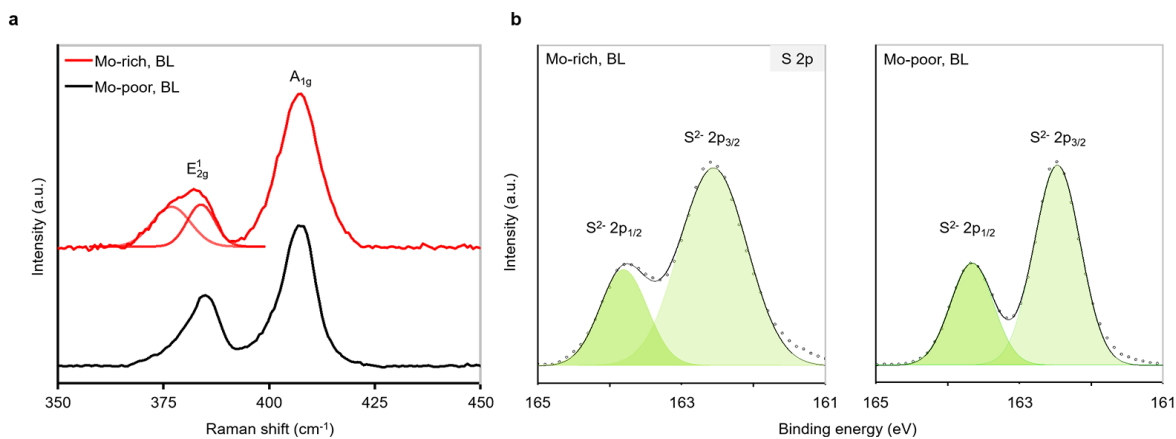

**Fig. S10. Raman spectra and XPS spectra of bilayer MoS$_2$**

(a) Raman spectra of transferred bilayer MoS$_2$ grown under Mo-rich (red) and Mo-poor (black) conditions. The bilayers were transferred onto fused silica substrates using a wet-transfer method. The sample grown under Mo-rich conditions exhibits splitting of the $E_{2g}^1$ peak, which differs from that observed in the sample grown under Mo-poor conditions. (b) XPS spectra of bilayer MoS$_2$ grown under Mo-rich (left) and Mo-poor (right) conditions, showing the S 2p.

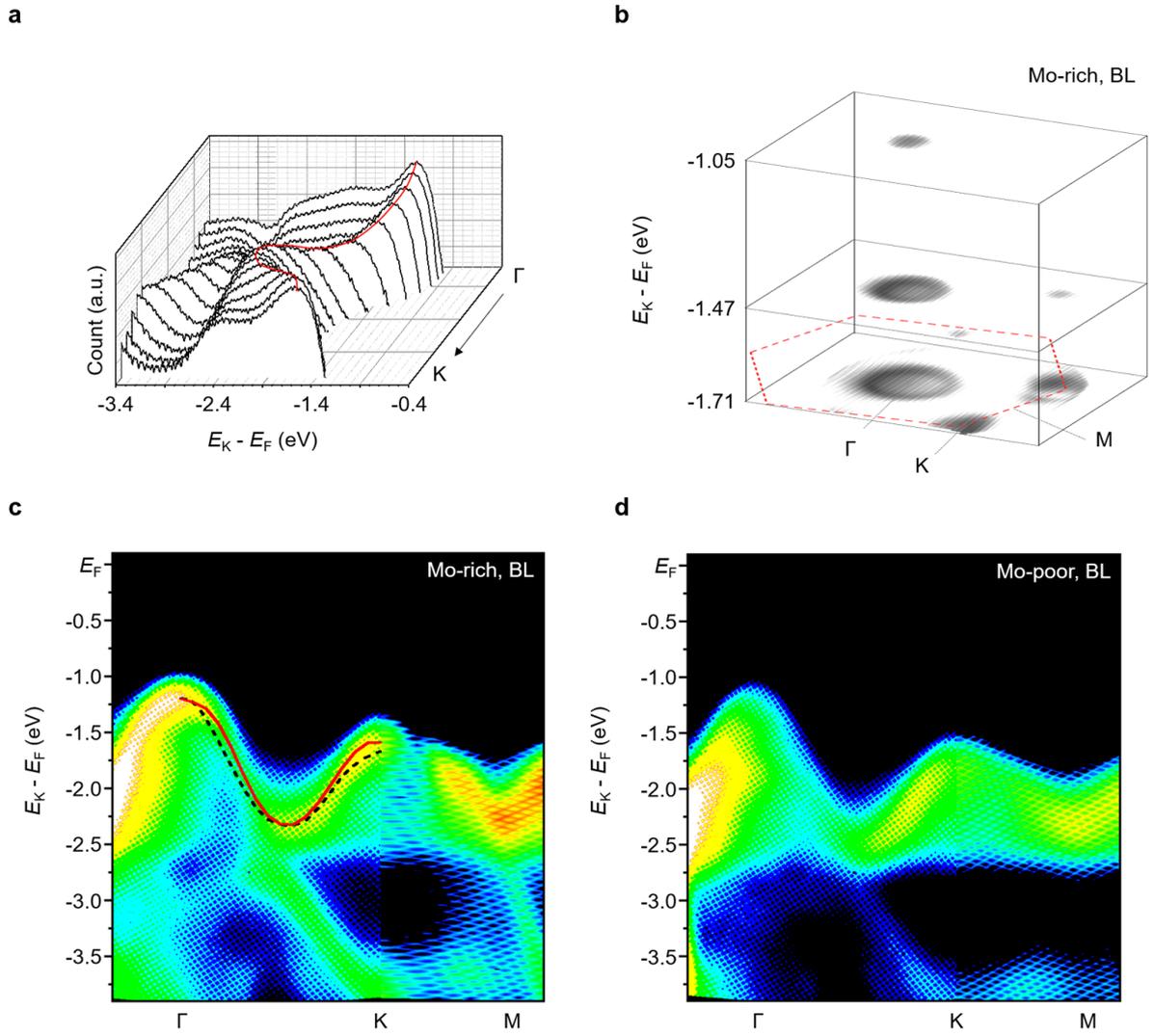

**Fig. S11. Band dispersions of bilayer MoS$_2$**

(a) Band dispersion along the Γ→K direction grown under Mo-rich conditions. The red line indicates the band maximum points. (b) ARPES mapping data of a bilayer MoS$_2$ sample grown under Mo-rich conditions. The z-axis represents $E_k - E_F$, and the XY-plane corresponds to momentum in the reciprocal lattice. The intensity on each plane visualizes the band distribution. (c), (d) Band dispersion of Γ-K-M direction obtained by ARPES which grown under Mo-rich and Mo-poor conditions, respectively. In the left panel, the red line marks the valence band maximum (VBM) of bilayer MoS$_2$ grown under Mo-rich conditions, while the black dotted line indicates the VBM from a reference[39].

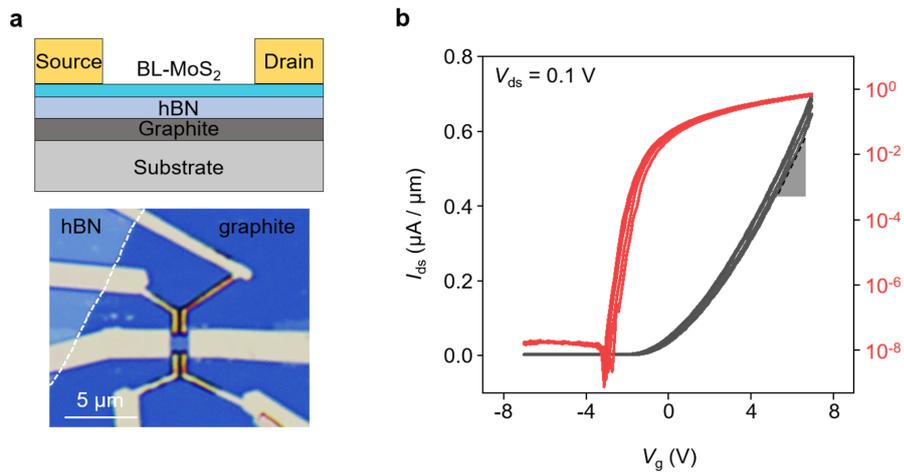

**Fig. S12. Device reliability during long-term operation of bilayer MoS$_2$ grown under Mo-rich conditions**

(a) Schematic of the device structure (top) and optical microscopy image of the fabricated field-effect transistor (FET) device (bottom). Scale bar, 5 μm ($L_{ch}$ = 1.35 μm, $W_{ch}$ = 2.2 μm). (b) Transconductance curve ($I_{ds}$ vs $V_g$) measured ten times over 5 hours at $V_{ds}$ = 0.1 V for a bilayer MoS$_2$ grown under Mo-rich conditions.

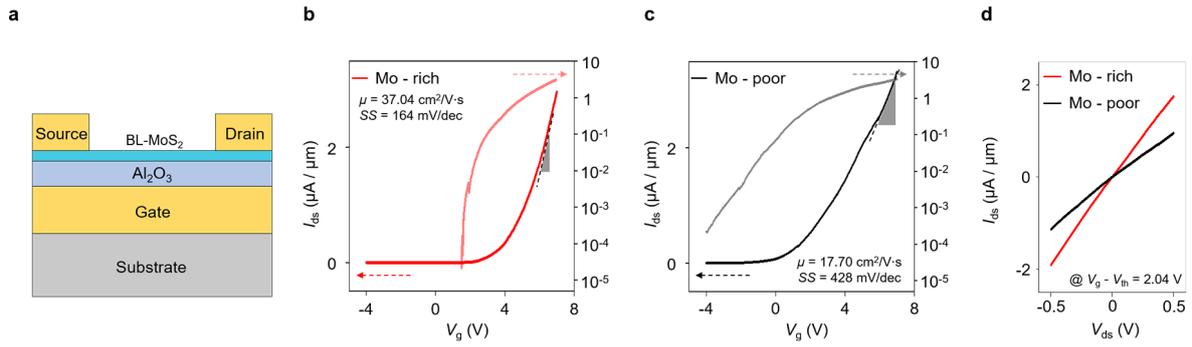

**Fig. S13. Electronic properties of bilayer MoS₂ grown under Mo-rich and Mo-poor conditions.**

(a) Schematic of the device structure (top) and optical microscopy image of the fabricated field-effect transistor (FET) device (bottom). Scale bar, 5 μm ($L_{ch}$ = 4.23 μm, $W_{ch}$ = 3.73 μm). (b) Transconductance curve ($I_{ds}$ vs $V_g$) measured at $V_{ds}$ = 1 V for bilayer MoS₂ grown under Mo-rich conditions. (c) Transconductance curve ($I_{ds}$ vs $V_g$) measured at $V_{ds}$ = 1 V for bilayer MoS₂ grown under Mo-poor conditions. (d) Output characteristics ($I_{ds}$ vs $V_{ds}$) at $V_g$ - $V_{th}$ = 2.04 V for devices fabricated with Mo-rich (red) and Mo-poor (black) bilayers.

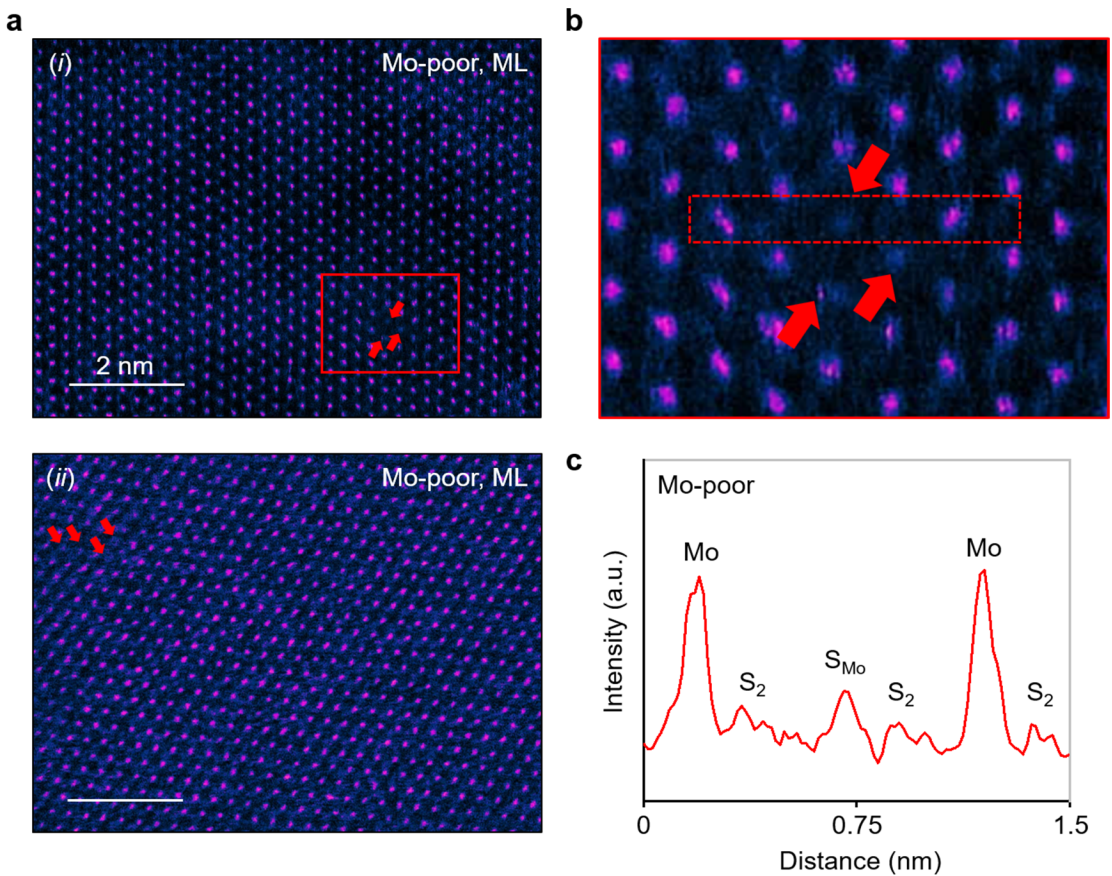

**Fig. S14. STEM image of a point defect in monolayer MoS$_2$ grown under Mo-poor conditions.**

(a) (*i*)–(*ii*) STEM images of monolayer MoS$_2$ grown under Mo-poor conditions. Red markers indicate point defects particularly sulfur substitutions at Mo sites (S$_{Mo}$). (b) High-resolution STEM image of the region highlighted in (a). (c) Intensity profile extracted from the boxed region in (b). Due to the mass contrast behavior of high-angle annular dark-field (HAADF) imaging, which scales with the square of the atomic number ($\sim Z^2$), a lower intensity at a Mo site suggests that a lighter S atom has substituted the Mo atom.

Bilayer MoS$_2$ channels grown under Mo-rich conditions exhibit superior electrical performance compared to those grown under Mo-poor conditions in the same device geometry, in terms of mobility, subthreshold swing (*SS*), and on/off current ratio (Fig. S13). Beyond the factors associated with stacking variations that degrade electrical performance in bilayer MoS$_2$ grown under Mo-poor conditions, as discussed in main manuscript, we also consider other plausible causes. Theoretically, growth under Mo-poor conditions could promote the formation of point defects such as Mo vacancies (V$_{Mo}$) and S substitutions (S$_{Mo}$)[1]. While atomic-resolution TEM images of bilayer MoS$_2$ grown under Mo-poor conditions reveal the presence of defect sites such as S$_{Mo}$, their density remains relatively low, estimated to be around $4.4\times10^{12}$ cm$^{-2}$, especially when compared to the Mo interstitial density in bilayer MoS$_2$ grown under Mo-rich conditions, suggesting limited effects on the electronic transports.

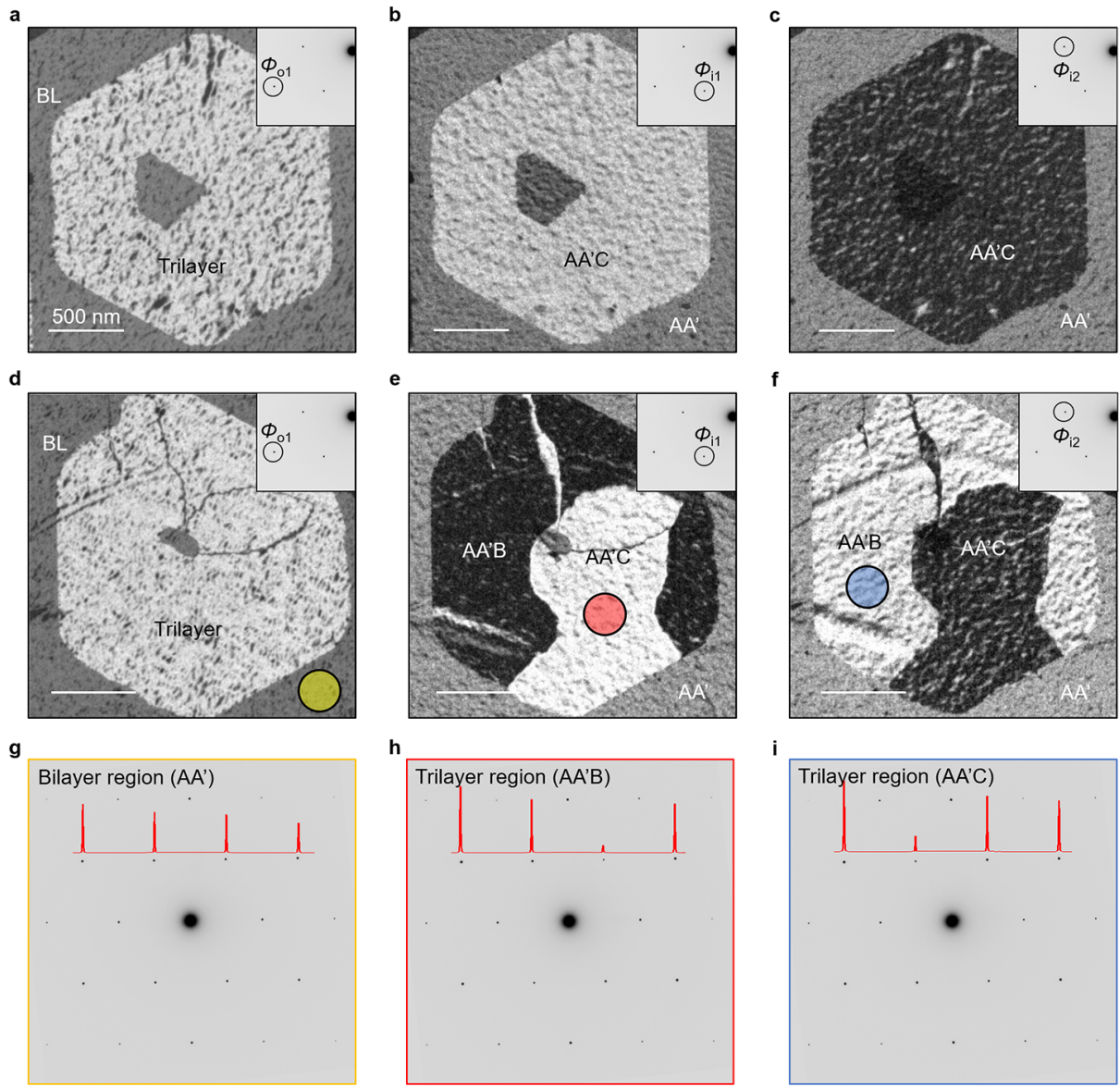

**Fig. S15. Dark-field TEM characterization of trilayer grains of MoS$_2$ grown under Mo-rich conditions.**

(a) Dark field Transmission Electron Microscopy (DF-TEM) image of trilayer regions obtained using an outer diffraction spot (upper right, $\Phi_{o1}$). (b, c) DF-TEM images acquired using inner diffraction spot ($\Phi_{i1}$ and $\Phi_{i2}$), respectively. (d) DF-TEM image obtained from another trilayer region using an outer diffraction spot (upper right, $\Phi_{o1}$). (e, f) Corresponding DF-TEM images acquired using an inner diffraction spot ($\Phi_{i1}$ and $\Phi_{i2}$), respectively. (g-i) Selected-area electron diffraction (SAED) patterns were obtained from the region indicated in the DF-TEM images. At trilayer, with one of the inner diffraction spots exhibiting an intensity approximately nine times lower than that of the outer spot. DF-TEM image (a, d) taken using the outer diffraction spot ($\Phi_{o1}$), exhibits local intensities proportional to the square of the number of layers ($n^2$), with distinct intensity contrasts observed in images taken from the inner diffraction spots ($\Phi_{i1}$ and $\Phi_{i2}$) depending on the diffraction direction (b-c, e-i). This observation clearly indicates that the third layer adopts a rhombohedral stacking order.

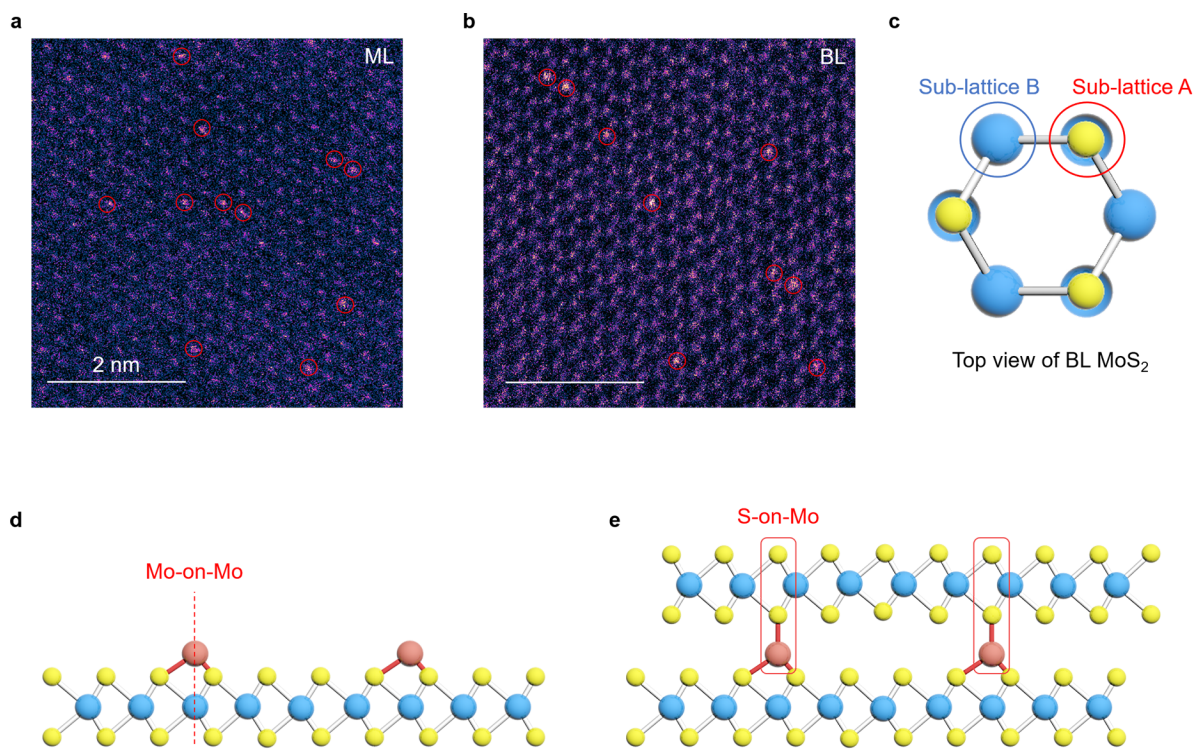

**Fig. S16. STEM analysis of Mo-adatoms and schematics of covalently bonded van der Waals interface**

(a) Scanning transmission electron microscopy (STEM) image of monolayer $MoS_2$ grown under Mo-rich conditions. Red circles indicate atomic sites exhibiting high scattering intensity. (b) STEM image of bilayer $MoS_2$ grown under Mo-rich conditions. Red circles indicate atomic sites with high scattering intensity. (c) Schematic illustration of the top-view sublattice structure of bilayer $MoS_2$. Sub-lattice A represents the S-on-Mo configuration, whereas sub-lattice B represents the Mo-on-S configuration. (d) Schematic of the Mo-on-Mo configuration, with Mo adatoms adsorbed on the Mo sites of monolayer $MoS_2$. (e) Schematic of the S-on-Mo configuration in bilayer $MoS_2$, where the 2H stacking order is induced by Mo interstitials. After bilayer growth, the state of the topmost $MoS_2$ layer changes from that of the original monolayer. At this stage, all covalently bonded Mo interstitials reside beneath the sulfur sites of the top layer, occupying one of the sublattices (b, c), and no additional Mo atoms adsorb on the surface, presumably due to local surface charge induced by the Mo interstitials. Consequently, the sulfur sites above Mo atoms (2e, red-highlighted) can serve as a new type of deterministically formed nucleation sites, replacing the Mo-on-Mo sites, and guide the rhombohedral stacking of the third layer.

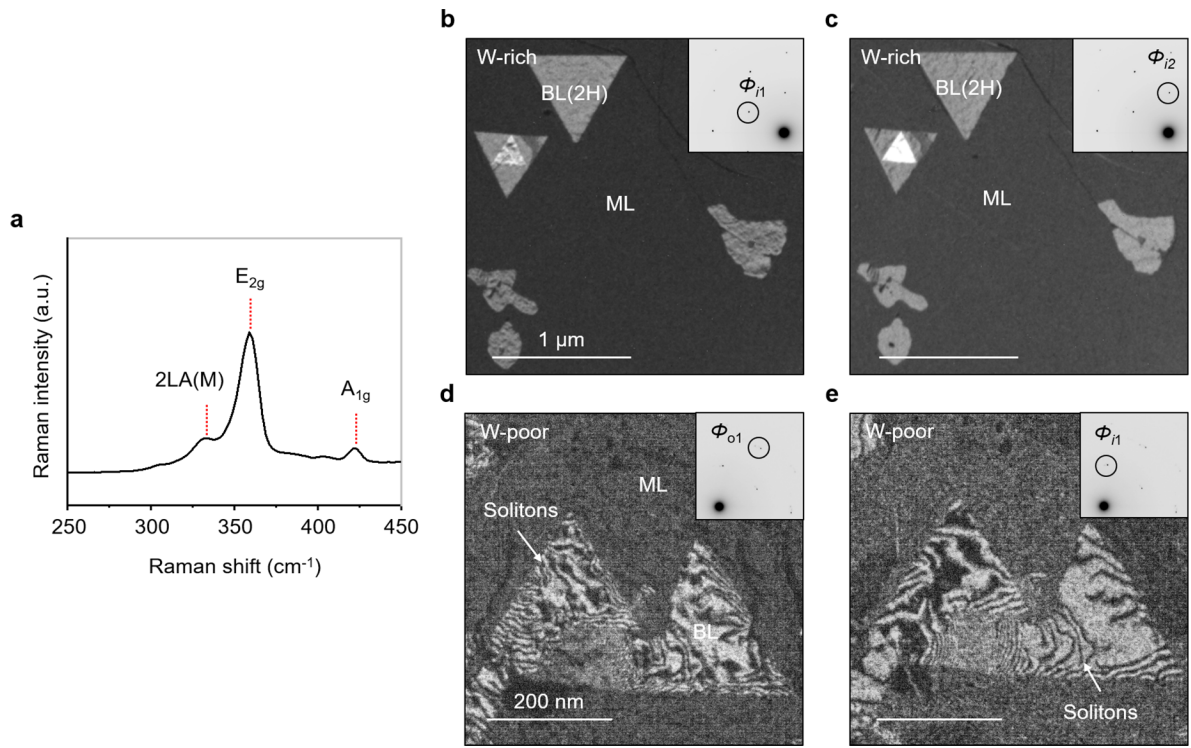

**Fig. S17. Robust bilayer WS$_2$ grown under W-rich conditions.**

(a) The Raman spectra of monolayer WS$_2$ shows an E$_{2g}$ peak at 358 cm$^{-1}$ and an A$_{1g}$ peak at 421 cm$^{-1}$. (b, c) Dark-field TEM images acquired at the inner diffraction spots ($\Phi_{i1}$ and $\Phi_{i2}$), respectively. The uniform intensity observed within the grain when selecting two different inner diffraction spots suggests that the structure has a 2H stacking order. In addition, the absence of strained solitons in the samples grown under W-rich conditions suggests that W atoms can also serve as interlayer linkers. (d, e) Dark-field TEM images acquired at the outer and inner diffraction spots ($\Phi_{o1}$ and $\Phi_{i1}$), respectively. The observed intensity variations within the bilayer grain correspond to strained solitons.